\documentclass{moriond}

\def\Journal#1#2#3#4{{#1} {\bf #2}, #3 (#4)}

\def\be{\begin{equation}}
\def\ee{\end{equation}}
\def\bea{\begin{eqnarray}}
\def\eea{\end{eqnarray}}

\newcommand{\Photo}{}

\begin{document}
\vspace*{4cm}
\title{NEUTRINO INTERACTIONS IN MICROBOONE}

\author{ MARCO DEL TUTTO \\
on behalf of the MicroBooNE collaboration}

\address{Department of Physics, University of Oxford,\\
Oxford OX1 3RH, United Kingdom}

\maketitle\abstracts{
MicroBooNE is a liquid-argon-based neutrino experiment, which began collecting data in Fermilab's Booster neutrino beam in October 2015. Physics goals of the experiment include probing the source of the anomalous excess of electron-like events in MiniBooNE. In addition to this, MicroBooNE is carrying out an extensive cross section physics program that will help to probe current theories on neutrino-nucleon interactions and nuclear effects. These proceedings summarise the status of MicroBooNE's neutrino cross section analyses.}

\section{Introduction}

The Micro Booster Neutrino Experiment (MicroBooNE) combines physics goals of short-baseline oscillations and neutrino cross section measurements with development goals to inform larger scale construction of Liquid Argon Time Projection Chambers (LArTPCs) for the long-baseline neutrino program. 
MicroBooNE is located on-axis in the 8 GeV Booster Neutrino Beam (BNB) line at Fermilab, 470 m downstream from the target.  
MicroBooNE finished commissioning in summer 2015 and has been collecting neutrino data since October 2015.

Many cross-section analyses are currently underway within the MicroBooNE collaboration, including studies of proton multiplicity, charged-current $\pi^0$ production, neutral-current channels, and more. Presented here are the first measurements of muon kinematic distributions, which are an intermediate step towards a $\nu_\mu$ charged-current inclusive cross section measurement. This analysis will provide a foundation for comparison to other experiments, and for the development of the tools required for more exclusive cross-section channels in future.

\section{The MicroBooNE Detector}

MicroBooNE is the first large (89 tons of active mass) LArTPC to operate in the United States, with dimensions of 10.4 (length) $\times$ 2.5 (width) $\times$ 2.3 (height) m.

\begin{figure}[t]
\centering
\begin{minipage}{20pc}
\includegraphics[width=20pc]{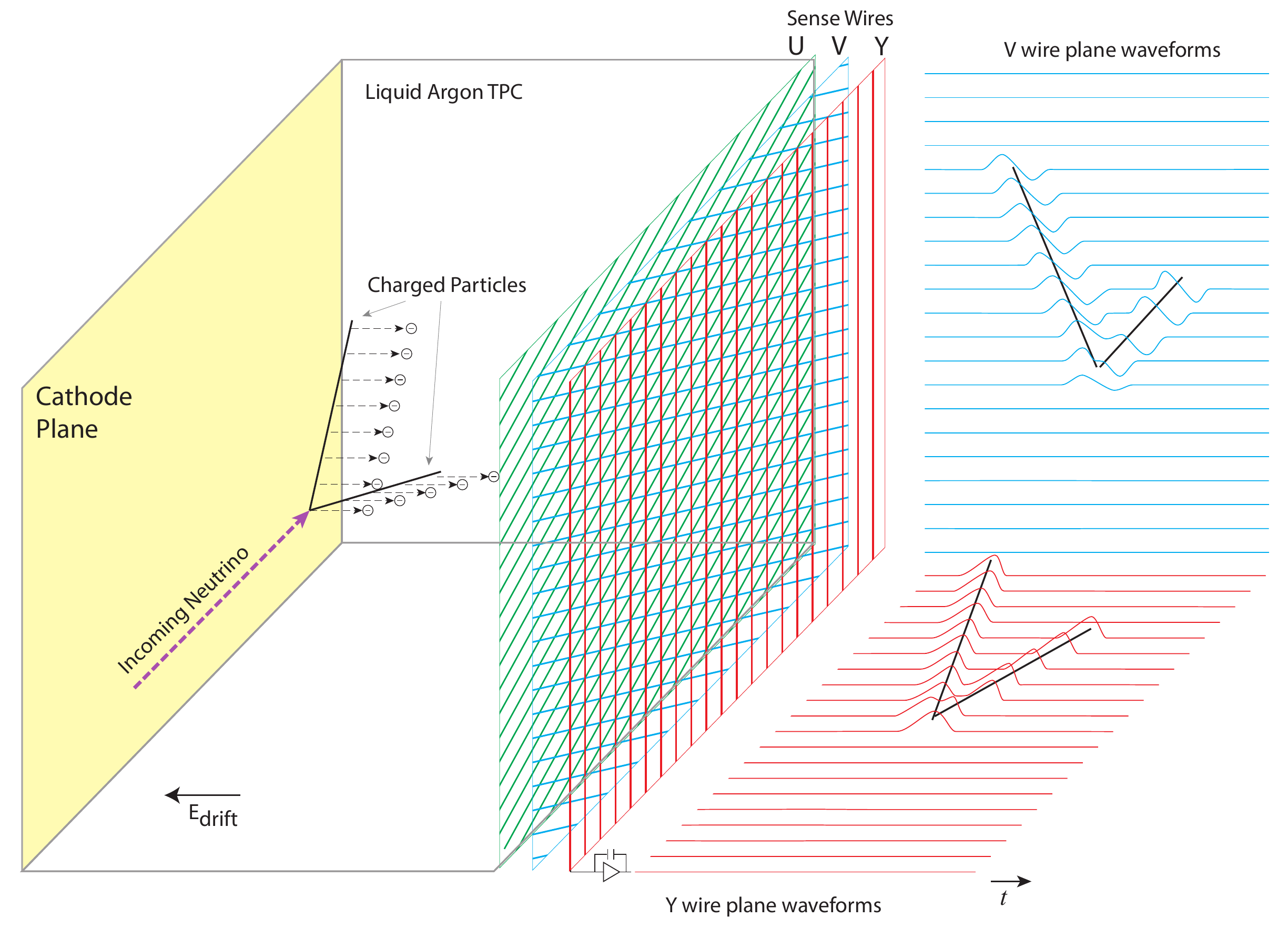}
%\caption{\label{label}Operational principle of the MicroBooNE LArTPC.}
\end{minipage}
\begin{minipage}{15pc}
\includegraphics[width=15pc]{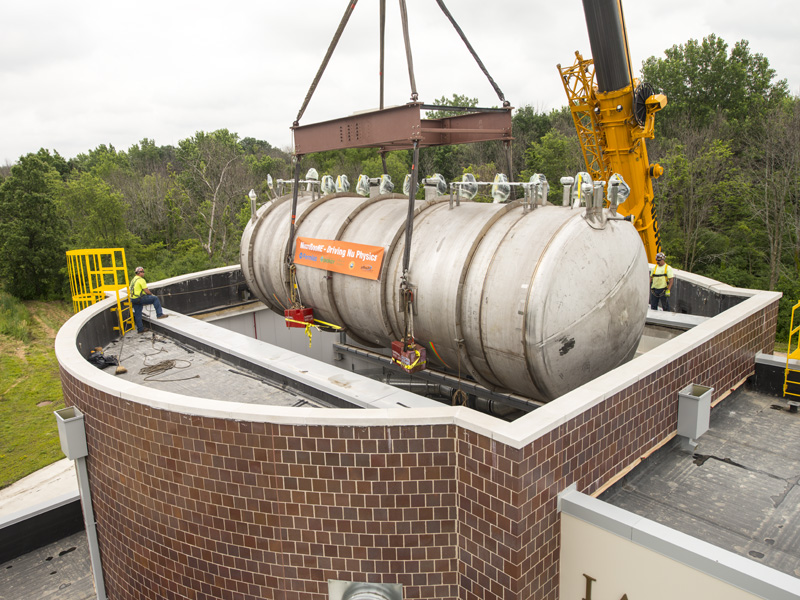}
%\caption{\label{label}A photograph of the cryostat, housing the TPC,  being lowered into the Liquid Argon Test Facility at Fermilab.}
\end{minipage}\hspace{2pc}
\caption{Operational principle of the MicroBooNE LArTPC (left). A photograph of the cryostat, housing the TPC,  being lowered into the Liquid Argon Test Facility at Fermilab (right).}
\label{fig:detector}
\end{figure}

In the MicroBooNE LArTPC detector,\cite{det} charged particles traversing a volume of highly-purified liquid argon leave trails of ionisation electrons in their wake and also create prompt vacuum ultraviolet scintillation photons, see Figure \ref{fig:detector}(left). The ionisation trails are transported practically undistorted over distances of the order of meters under the influence of a uniform electric field until they reach anode planes located along one side of the active volume. The anode planes are composed of wires that receive the signals induced by the ionisation electrons drifting towards them. 
The charged particle trajectory reconstruction is derived from the known wire positions within the anode planes and the drift time of the ionisation. The drift time is the difference between the arrival times of ionisation signals on the wires and the $t_0$ time the interaction took place in the detector which is established by a trigger provided by the light collection system: 32 photo-multipliers (PMTs) located on one side of the TPC volume.

MicroBooNE is a surface detector and is therefore subject to constant bombardment by cosmic radiation (or cosmics) from the atmosphere. Quantitatively, this amounts to $\sim$5 kHz, or $\sim$20-30 cosmics per MicroBooNE's 4.8 ms drift time window. However, neutrinos are not delivered continuously but in distinct spills. The BNB spill (1.6 $\mu$s) is much smaller than the drift time, and a natural way to separate cosmics from neutrino interactions is to compare the time of the activity in the detector with the spill trigger time window. The 32 PMTs light collection system signals are then vital in distinguishing detector activity that is in-time with the beam from that which is out-of-time. Spatial information can also be inferred from the light collection signals, further aiding in the eventual reconstruction of the activity occurring inside the detector.

\begin{figure}[!ht]
\centering
\begin{minipage}{13.9pc}
\includegraphics[width=13.9pc]{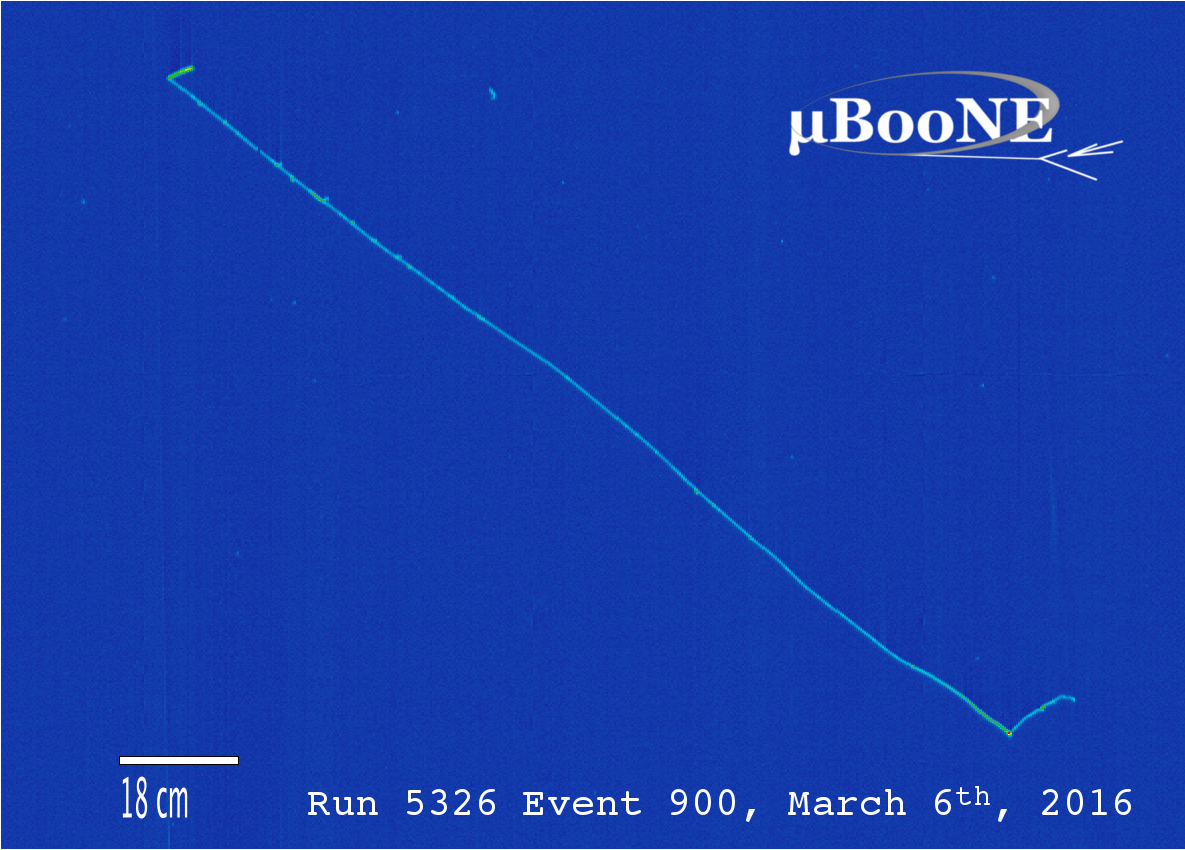}
%\caption{}
\end{minipage}
\begin{minipage}{15.2pc}
\includegraphics[width=15.2pc]{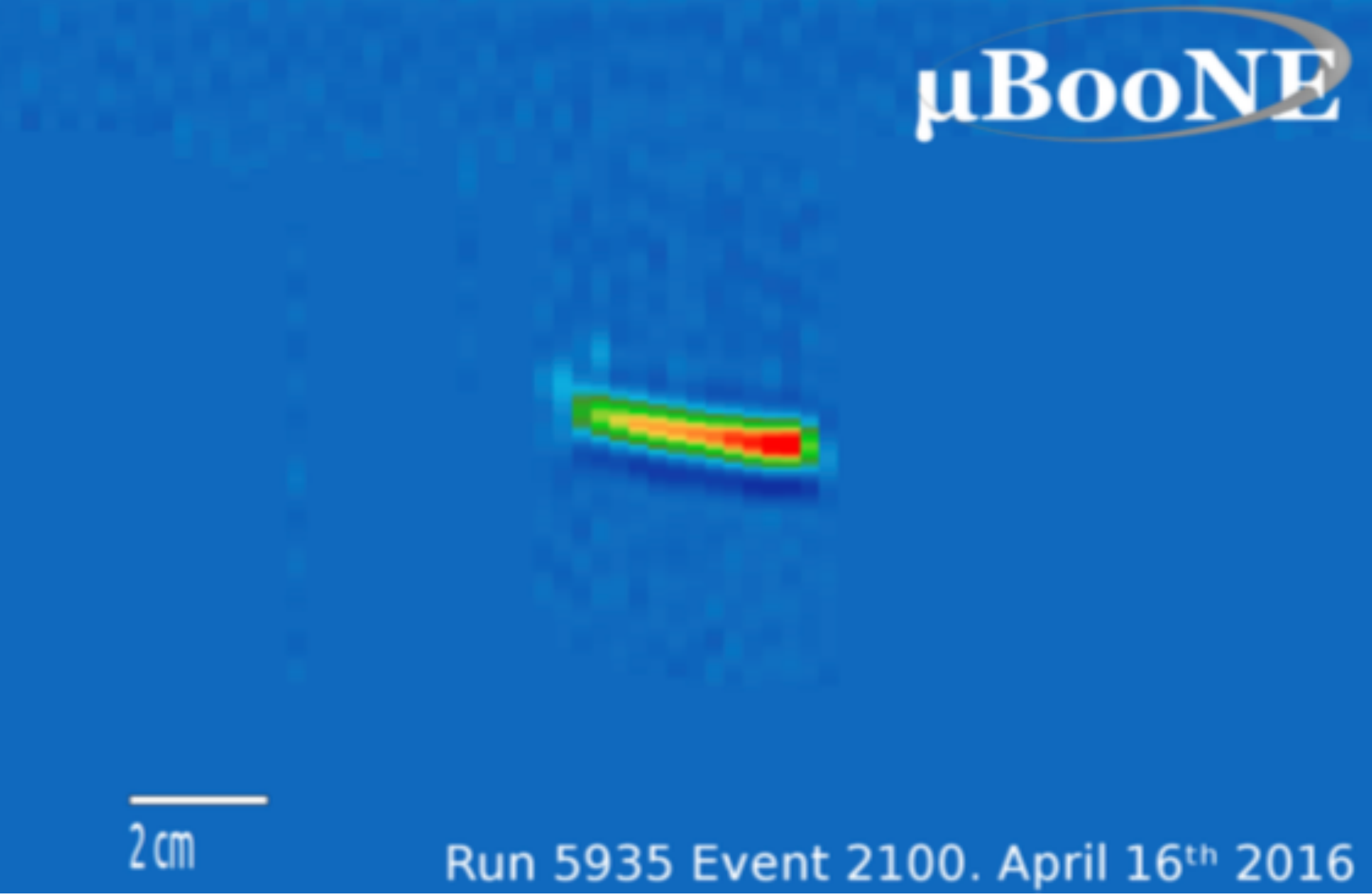}
%\caption{}
\end{minipage}\hspace{2pc}%
\caption{Event displays showing neutrino interactions in MicroBooNE. Left is a CC and right a NC candidate event. These are collection plane images. Neutrino beam is coming from the left. Colour indicates amount of deposited charge on the wires. On the left, the long track is a muon candidate, the short track on the upper left corner is a proton candidate, while the track at the end of the muon is a Michel electron candidate. On the left, the only short visible track is a proton candidate.}
\label{fig:evd-cc-nc}
\end{figure}

\section{Neutrino Interactions}

One of the main physics goals of MicroBooNE is to perform high-statistics precision measurements of $\nu$-Ar interactions in the 1 GeV range. Cross section measurements allow us to understand the neutrino interaction models and nuclear effect which are of fundamental importance for future oscillation experiments.

MicroBooNE, due to its large size, is collecting large statistics and analyses are expected to be only systematics limited. MicroBooNE will collect around 170,000 $\nu_\mu$ CC inclusive interactions in 3 years, opening a great opportunity to study nuclear effects in argon. 

\begin{figure}[!ht]
\centering
\begin{minipage}{15.8pc}
\includegraphics[width=15.8pc]{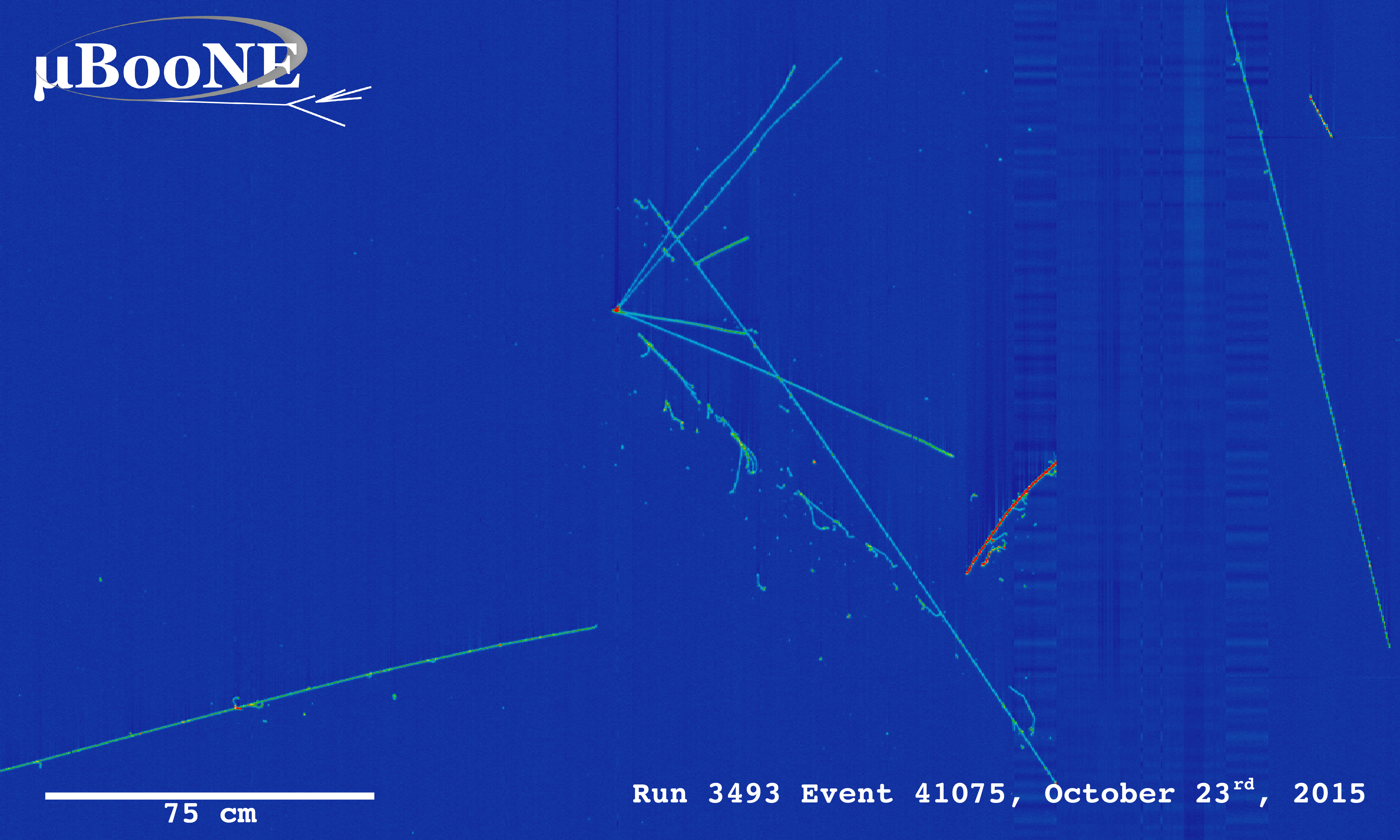}
%\caption{}
\end{minipage}
\begin{minipage}{13.1pc}
\includegraphics[width=13.1pc]{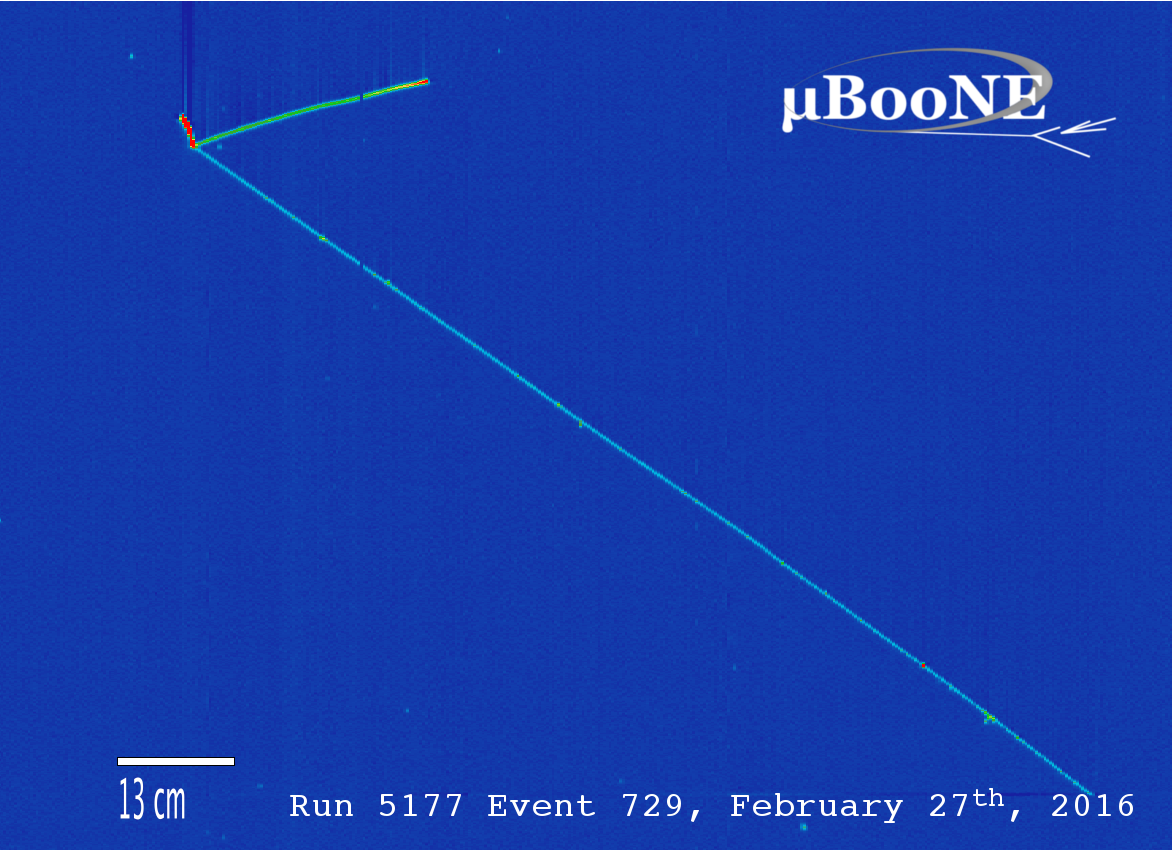}
%\caption{}
\end{minipage}\hspace{2pc}%
\caption{Event displays showing neutrino interactions in MicroBooNE. Left is a CC-multitrack and right a CC-$1\pi$ candidate event. These are collection plane images. Neutrino beam is coming from the left. Colour indicates amount of deposited charge on the wires.}
\label{fig:evd}
\end{figure}

MicroBooNE is able to detect both neutral-current (NC) and charged-current (CC) events. Figure \ref{fig:evd-cc-nc} shows a CC event candidate (left) and a NC one\cite{nc} (right) selected from data by automated event selections. Figure \ref{fig:evd} shows two neutrino candidate events in MicroBooNE: a CC-multitrack (left) and a CC-1$\pi$ (right).

\section{MicroBooNE Charged-Current Inclusive Event Selection}

In order to perform high-statistics physics measurements, MicroBooNE requires automated selection algorithms, which in turn require automated reconstruction as input. 

MicroBooNE has developed two complementary (preliminary) selections to select charged-current muon neutrino interactions in the liquid argon. Both are fully automated, cut-based selections. The results presented here are those given by the second selection, but the full details of both selections can be found in a MicroBooNE public note.\cite{numu} 
Here is a brief description of the event selection used. The event must have no less than one flash, which is higher than 50 photo-electrons, inside of the beam spill window, indicating the possible presence of a neutrino interaction. If there is at least one track with a closest approach distance of 70 cm from the beam flash center in the Z direction, the event is selected.

The selection then takes different paths depending on the number of tracks produced from each reconstructed vertex. If only one track is produced from a reconstructed vertex, the track is selected only if it is fully contained and passes a minimum track length requirement. If two tracks are produced from the vertex, they are selected only if the vertex has not been tagged as a point where a cosmic stopping muon decays to a Michel electron. If more than two tracks are produced, the event is automatically selected. In the cases where there is more than one track, the longest reconstructed track associated with the selected vertex in the surviving event is tagged as the muon candidate.

\begin{figure}[!ht]
\centering
\includegraphics[width=17pc]{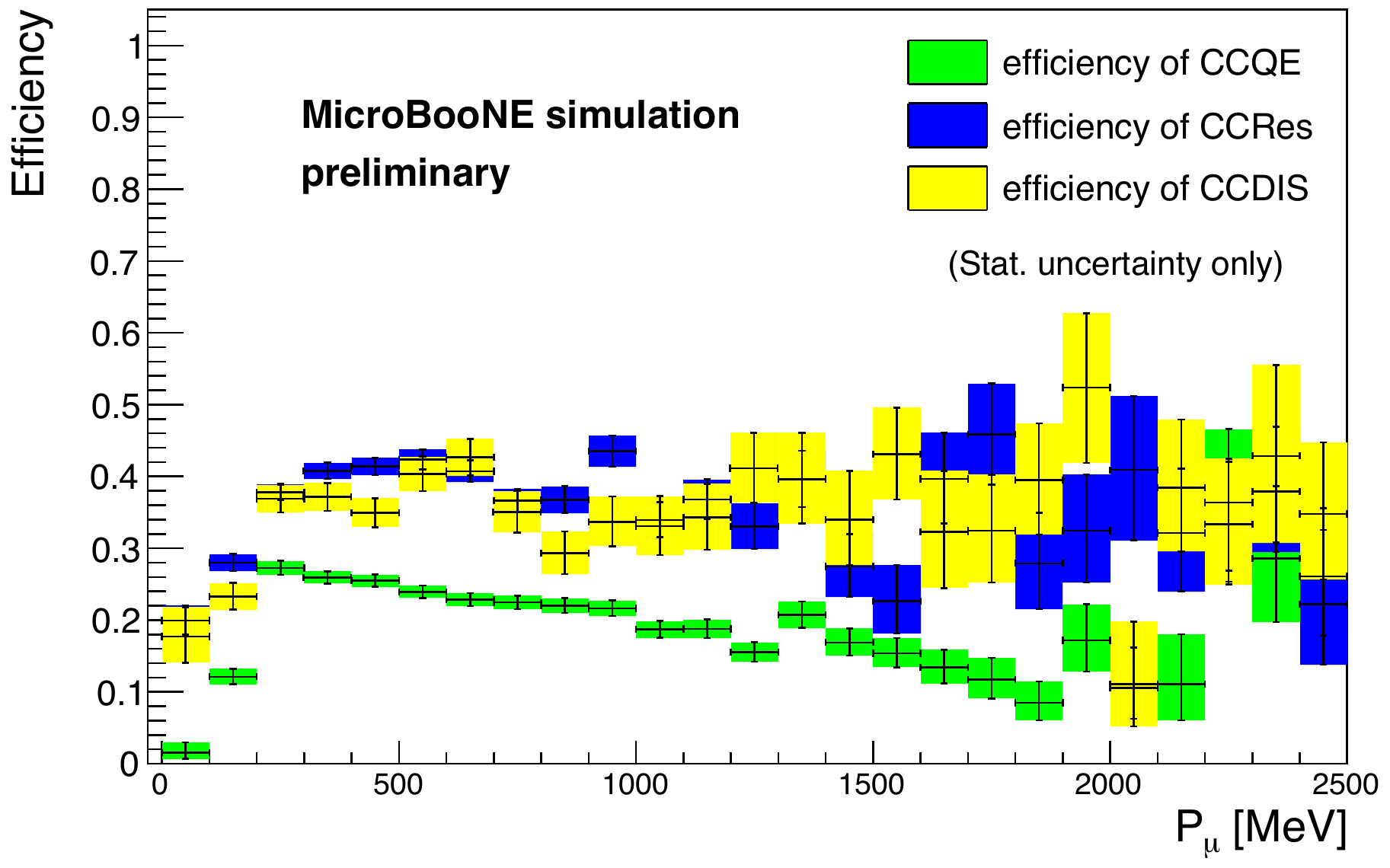}
\caption{The efficiency of the selected events for the processes quasi-elastic (QE), resonant (RES), and deep-inelastic (DIS). The bands represent the statistical uncertainty on the simulation only. The distribution is shown as a function of the true muon momentum. The rise of the efficiency between 0 and 0.5 GeV is caused by the minimum track length cut, which directly translates into a cut on the muon momentum. There is no decrease in the efficiency for higher momentum RES and DIS events, because there is no containment requirement for events with more than one track produced from the vertex.}
\label{fig:eff}
\end{figure}

These cuts result in an overall efficiency $\times$ acceptance of 30\%, and an overall purity of 65\%, the main background coming from cosmic rays. The efficiency is shown in Figure \ref{fig:eff} as a function of the muon momentum.

% FIGURE - Efficiency

\begin{figure}[!ht]
\centering
\begin{minipage}{18.5pc}
\includegraphics[width=18.5pc]{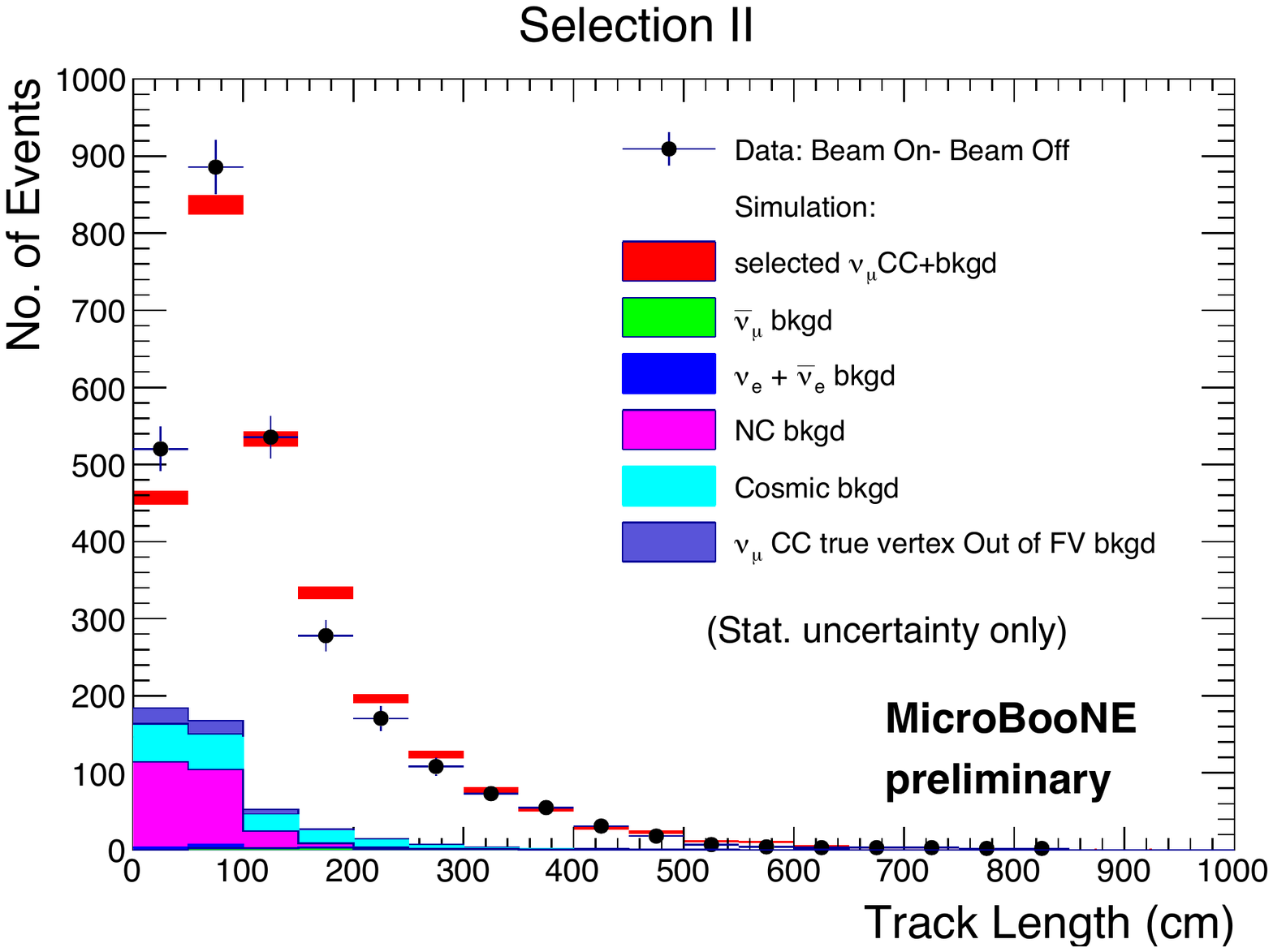}
%\caption{}
\end{minipage}
\begin{minipage}{18.5pc}
\includegraphics[width=18.5pc]{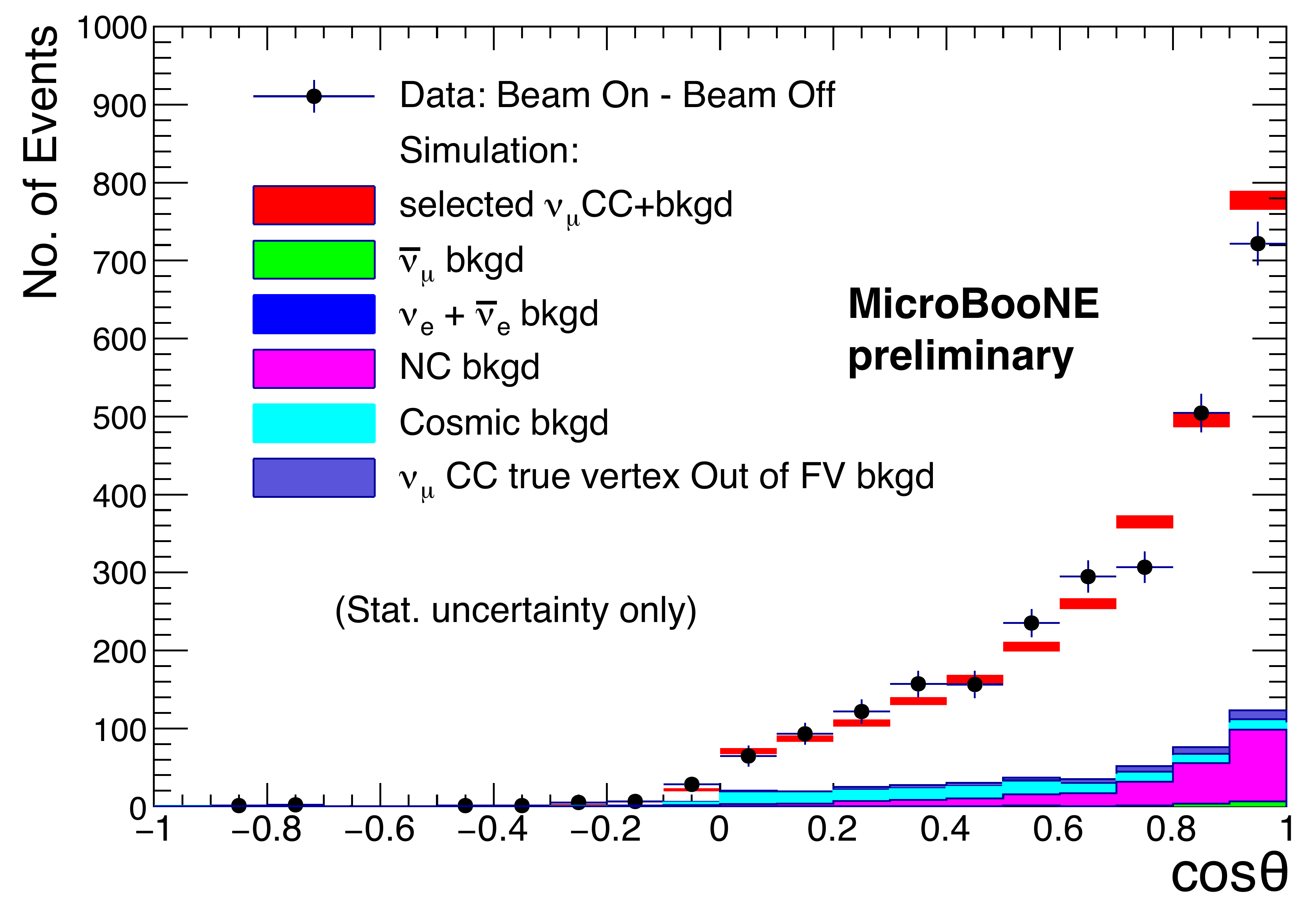}
%\caption{}
\end{minipage}\hspace{2pc}%
\caption{The black data points symbolise on-beam minus off-beam data with statistical error bars. The on-beam data sample contains events where no neutrino interacted in the detector (a purely cosmic event) and events where a neutrino interaction is also present. Off-beam data are taken outside of the beam spill window and contain only cosmic events. The subtraction of off-beam from on-beam data subtracts the background of events of purely cosmic events. The red shaded histogram represents the Monte Carlo, with the bands representing the statistical uncertainty only. The backgrounds contained in the red are additionally shown in different colors corresponding to different physics processes. Backgrounds are stacked. The simulation is normalized to the same number of total events as the data.}
\label{fig:dist}
\end{figure}

Applying this selection to a subset of data equivalent to $5\times 10^{19}$ protons on target, and performing a background subtraction of the surviving cosmics outside the beam window using beam-off data, we observe the kinematic distributions shown in Figure \ref{fig:dist}. The Figure shows the distribution of the muon momentum and the muon track's angle with respect to the beam direction ($\theta$). These distributions are presented with statistical uncertainties only, and with Monte Carlo distributions scaled to the same number of events as the data; the assessment of the systematic uncertainties is still ongoing.

A third and final CC inclusive selection, which combines and improves upon these two selections, is currently underway.

% FIGURE - Event distributions

\section{Conclusions}

MicroBooNE has made its first kinematic measurements of neutrino interactions in a muon neutrino beam. A charged-current inclusive cross-section will soon follow. These measurements demonstrate that a large-scale liquid argon TPC can make excellent measurements of neutrino interactions, and provide the foundation for a much more wide-ranging programme of cross-section measurements to come.


\section*{References}
\begin{thebibliography}{99}
\bibitem{det}MicroBooNE Collaboration, \Journal{JINST}{12}{P02017}{2017}.

\bibitem{numu}MicroBooNE-Public-Note-1010 (2016) \url{http://microboone.fnal.gov/public-notes/}.

\bibitem{nc}MicroBooNE-Public-Note-1025 (2016) \url{http://microboone.fnal.gov/public-notes/}.

\end{thebibliography}
\end{document}